\documentclass[aps,prl,reprint,floatfix,groupedaddress,superscriptaddress,showpacs]{revtex4-1}

\usepackage{amsmath}
\usepackage{graphicx}
\usepackage{natbib}

\begin{document}

%%%%%%%%%%%%%%%%%%%%%%%%%%%%%%%%%%%%%%%%%%%%%%%%%%%%%%%%%%%%%%%%%%%%%%%%%%%%%%%%%%%%%%%%%%%%%%%%%%%%%%%%%%%%
% Title
%%%%%%%%%%%%%%%%%%%%%%%%%%%%%%%%%%%%%%%%%%%%%%%%%%%%%%%%%%%%%%%%%%%%%%%%%%%%%%%%%%%%%%%%%%%%%%%%%%%%%%%%%%%%

\title{Magnetic soft modes in the locally distorted triangular antiferromagnet $\alpha$-CaCr$_2$O$_4$}\

\author{S. Toth}
\email{sandor.toth@helmholtz-berlin.de}
\affiliation{Helmholtz Zentrum Berlin f\"{u}r Materialien und Energie, Hahn-Meitner-Platz 1, D-14109 Berlin, Germany}
\affiliation{Institut f\"{u}r Festk\"{o}rperphysik, Technische Universit\"{a}t Berlin, Hardenbergstr.\ 36, D-10623 Berlin, Germany}

\author{B. Lake}
\affiliation{Helmholtz Zentrum Berlin f\"{u}r Materialien und Energie, Hahn-Meitner-Platz 1, D-14109 Berlin, Germany}
\affiliation{Institut f\"{u}r Festk\"{o}rperphysik, Technische Universit\"{a}t Berlin, Hardenbergstr.\ 36, D-10623 Berlin, Germany}

\author{K. Hradil}
\affiliation{Forschungs-Neutronenquelle Heinz Maier-Leibnitz, D-85747 Garching, Germany}

\author{T. Guidi}
\affiliation{ISIS Facility, STFC Rutherford Appleton Laboratory, Chilton, Didcot, Oxon OX11 0QX, United Kingdom}

\author{K. C. Rule}
\affiliation{Helmholtz Zentrum Berlin f\"{u}r Materialien und Energie, Hahn-Meitner-Platz 1, D-14109 Berlin, Germany}

\author{M. B. Stone}
\affiliation{Neutron Sciences Directorate, Quantum Condensed Matter Division, Oak Ridge National Laboratory, Oak Ridge, Tennessee 37831, USA}

\author{A. T. M. N. Islam}
\affiliation{Helmholtz Zentrum Berlin f\"{u}r Materialien und Energie, Hahn-Meitner-Platz 1, D-14109 Berlin, Germany}

\date{\textrm{\today}}

\pacs{61.05.F-, 75.10.Hk, 75.30.Ds, 75.30.Et}
% 61.05.F- Neutron diffraction and scattering 
% 75.10.Hk Classical spin models
% 75.30.Ds Spin waves
% 75.30.Et Exchange and superexchange interactions

\begin{abstract}
	In this paper we explore the phase diagram and excitations of a distorted triangular lattice antiferromagnet. The unique two-dimensional distortion considered here is very different from the 'isosceles'-type distortion that has been extensively investigated. We show that it is able to stabilize a 120$^\circ$ spin structure for a large range of exchange interaction values, while new structures are found for extreme distortions. A physical realization of this model is $\alpha$-CaCr$_2$O$_4$ which has 120$^\circ$ structure but lies very close to the phase boundary. This is verified by inelastic neutron scattering which reveals unusual roton-like minima at reciprocal space points different from those corresponding to the magnetic order.
\end{abstract}

\maketitle

%%%%%%%%%%%%%%%%%%%%%%%%%%%%%%%%%%%%%%%%%%%%%%%%%%%%%%%%%%%%%%%%%%%%%%%%%%%%%%%%%%%%%%%%%%%%%%%%%%%%%%%%%%%%
% Introduction
%%%%%%%%%%%%%%%%%%%%%%%%%%%%%%%%%%%%%%%%%%%%%%%%%%%%%%%%%%%%%%%%%%%%%%%%%%%%%%%%%%%%%%%%%%%%%%%%%%%%%%%%%%%%

In frustrated antiferromagnets (AF) the magnetic interactions compete and it is impossible to satisfy them all  simultaneously. This competition causes degenerate ground states to occur, and the system fluctuates between them suppressing long-range magnetic order to temperatures well below the Curie-Weiss temperature \cite{ramirez2001}. Most frustrated magnets do eventually order and in the case of isotropic or Heisenberg interactions a compromise is often reached where the magnetic interactions are all partially satisfied by a non-collinear arrangement of the magnetic moments. 

The triangular lattice Heisenberg antiferromagnet (TLHAF) is one of the simplest examples of an extended frustrated magnetic system, however open questions  remain concerning its physical properties. In the case of spin-1/2, the ground state is expected to show spin-liquid behavior and possibly form a resonating valance bond (RVB) state \cite{Moessner2001}. For spin-1, both a RVB state and a spin nematic phase have been proposed \cite{Moessner2001,Tsunetsugu2006}. Systems with larger spin values order in a helical structure characterized by a magnetic ordering wavevector of 1/3 where the angle between nearest neighbor spins is 120$^\circ$.

Linear spin-wave theory (LSWT) for the TLHAF predicts one sinusoidal, doubly degenerate mode. However the excitation spectrum for spin-1/2 is expected to strongly deviate from LSWT; downward renormalization (of up to 33\%) and damping of excitations are expected over large regions of the Brillouin-zone \cite{Starykh2006,Zheng2006,Chernyshev2009}. A reduced effect is expected for higher spin values. Introduction of antiferromagnetic next-nearest-neighbor interactions also modifies the excitations and drives the magnetic structure from helical to collinear when the ratio of next-nearest to nearest neighbor interactions is greater than the critical value of 0.125 \cite{Wheeler2009a}. In the collinear phase, quantum fluctuations lift the classical degeneracy and select collinear stripe order via the mechanism of order by disorder \cite{Chubukov1992}, they also significantly alter the spin-wave excitation spectra around certain regions of the Brillouin zone \cite{Wheeler2009a}.
	\begin{figure}[!htb]
		\centering
		\includegraphics[width= 86 mm]{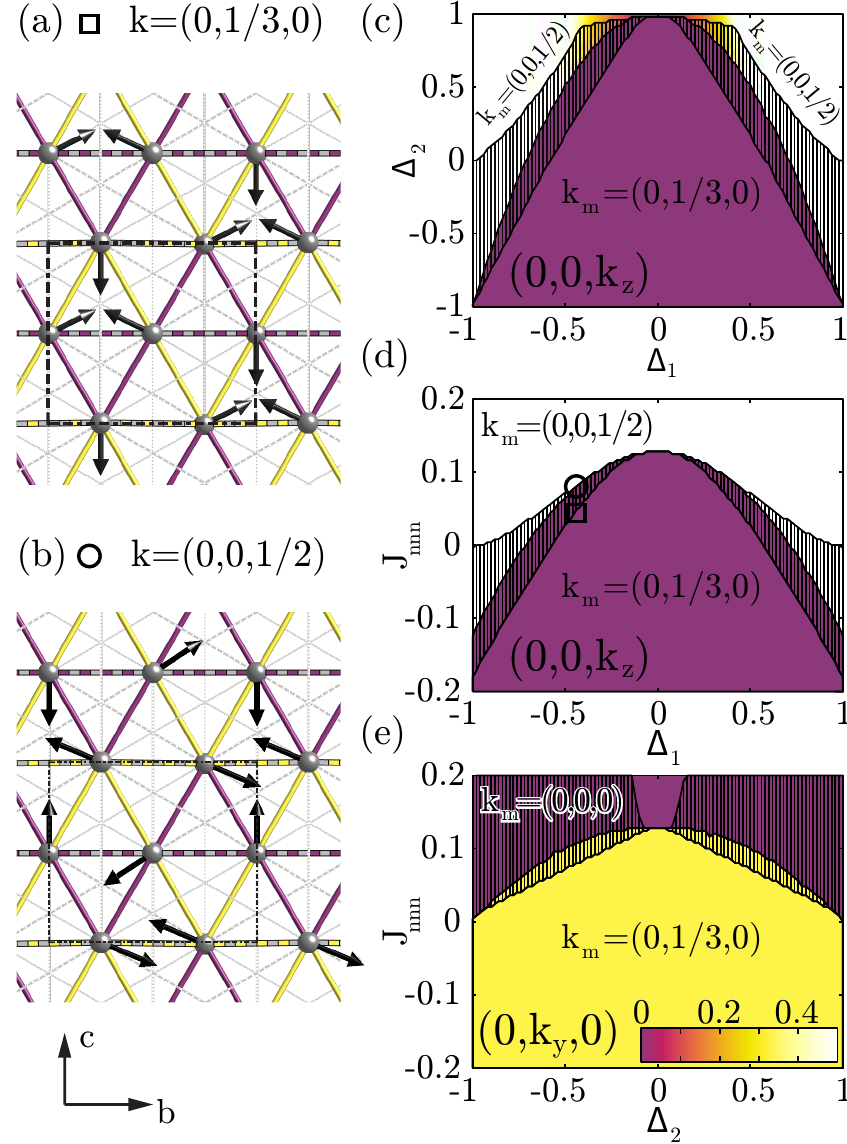}
		\caption{(color online) (a-b) Ordered magnetic structure of single triangular layer of Cr$^{3+}$ ions, arrows represent the magnetic moments and spins are rotated to the $bc$ plane for better visibility. (a) and (b) give that structure at the position of the square ($k_\text{m} = (0,1/3,0)$) and circle ($k_\text{m} = (0,0,1/2)$) shown on the phase diagram (d) respectively. The solid lilac and yellow lines represent $J_\text{zz1}$ and $J_\text{zz2}$, the dashed lines are $J_\text{ch1}$ and $J_\text{ch2}$ respectively. Thin gray lines depict next-nearest-neighbor interactions.(c-e) Magnetic ordering wavevector as a function of $\Delta_1$, $\Delta_2$ and $J_\text{nnn}$ for $\Delta_3 = 0$, the colors indicate the size of the components of $k_\text{m}$, in the shaded region the magnetic ground state has a multi-$k$ structure.}
		\label{fig:struct}
	\end{figure}
	
	Most physical realizations of triangular lattices are distorted. The distortion usually constitutes a shortening of the nearest neighbor distances along one of the three equivalent directions, resulting in one of the exchange constants being larger than the other two. This distortion partially lifts the frustration and promotes the formation of chains, with frustrated interchain coupling as in\ Cs$_{2}$CuCl$_{4}$ \cite{Coldea1996} and CuCrO$_{2}$ \cite{Poienar2010}. The resulting ordering wavevector is intermediate between the undistorted value of 1/3 and the AF chain value of 1/2. 

	In this paper, we investigate the excitation spectrum of $\alpha$-CaCr$_2$O$_4$ which is another distorted triangular lattice antiferromagnet but where the distortion is very different from those investigated previously. Furthermore, in addition to interactions between nearest neighbor magnetic ions there are also substantial next-nearest-neighbor interactions. This system allows examination of the effects of competing nearest-neighbor and next-nearest-neighbor interactions on a distorted triangular lattice where the distortion places the system close to a magnetic phase boundary.
	
 $\alpha$-CaCr$_2$O$_4$ belongs to the orthorhombic $Pmmn$ space group and has a distorted delafossite structure \cite{Pausch1974}. The magnetic Cr$^{3+}$ ions lie in an octahedral environment and are characterized by a half-filled t$_\text{2g}$ shell giving rise to spin-3/2 and quenched orbital angular momentum which ensure that the magnetic interactions are isotropic. They lie at two inequivalent crystallographic positions and form distorted triangular layers stacked along the $a$ axis. There are two layers and eight magnetic ions per unit cell.  Together with $\alpha$-SrCr$_2$O$_4$ \cite{Dutton11}, this compound shows a unique distortion, see Fig.\ \ref{fig:struct}(a). Although the Cr$^{3+}$ ions are locally displaced from the ideal triangular positions, the ratio of the $b$ and $c$ lattice parameters remain close to the triangular $2/\sqrt{3}$ value. The four inequivalent nearest neighbor  Cr$^{3+}$--Cr$^{3+}$ distances range from 2.889 \AA\ to 2.939 \AA\ and form two zig-zag and two straight chains. The short Cr$^{3+}$--Cr$^{3+}$ bonds and the almost 90$^\circ$ Cr$^{3+}$--O$^{2-}$--Cr$^{3+}$ angles mean that the dominant exchange mechanism is direct exchange as in many other Cr$^{3+}$ delafossite compounds \cite{Poienar2010, Frontzek2011, Kan2009, Hsieh2008a, Moreno2004, Delmas1978, Meisenheimer1961, Takatsu2009}. The Curie-Weiss temperature of -564(4) K, yields an average nearest neighbor exchange constant of $J_\text{nn}=6.48$ meV \cite{Chapon2011,Toth2011}. However the sensitive distance dependence of direct exchange would suggest large differences between the inequivalent nearest-neighbor interaction strengths. The interplane distances are much greater ($\sim$5.524 \AA) and are therefore expected to be mediated by much weaker superexchange interactions.

	$\alpha$-CaCr$_2$O$_4$ orders magnetically below $T_\text{N}= 42.6$ K \cite{Chapon2011,Toth2011}. The ordered state is the same as for the undistorted TLHAF with angles of 120$^\circ$ between nearest neighbors and spins lying in the $ac$ plane, (Fig.\ \ref{fig:struct}(a)). In ref.\ \onlinecite{Toth2011}, a classical model was developed which showed that this 120$^\circ$ structure is stable over a large range of the exchange constant values and provided constraints on their relative sizes. In this paper, we show that despite the simplicity of the magnetic structure, the magnetic excitations of $\alpha$-CaCr$_2$O$_4$ are highly complex. Furthermore, a significant softening of the spin-waves was found around wavevectors such as (0,1/2,0) and (0,0,1) placing it close to an instability away from the simple 120$^\circ$ structure towards the formation of a complex multi-$k$ structure. 

%%%%%%%%%%%%%%%%%%%%%%%%%%%%%%%%%%%%%%%%%%%%%%%%%%%%%%%%%%%%%%%%%%%%%%%%%%%%%%%%%%%%%%%%%%%%%%%%%%%%%%%%%%%%
% Experimental details
%%%%%%%%%%%%%%%%%%%%%%%%%%%%%%%%%%%%%%%%%%%%%%%%%%%%%%%%%%%%%%%%%%%%%%%%%%%%%%%%%%%%%%%%%%%%%%%%%%%%%%%%%%%%

	Both polycrystalline and single-crystal samples of $\alpha$-CaCr$_2$O$_4$ were measured \cite{nazmul3000}. The polycrystalline sample has a mass of 3.5 g; the single-crystal is 340 mg and consists of three crystallographic twins rotated by 60$^\circ$ with respect to each other about the shared $a$-axis. 
	\begin{figure}[!htb]
		\centering
		\includegraphics[width= 86 mm]{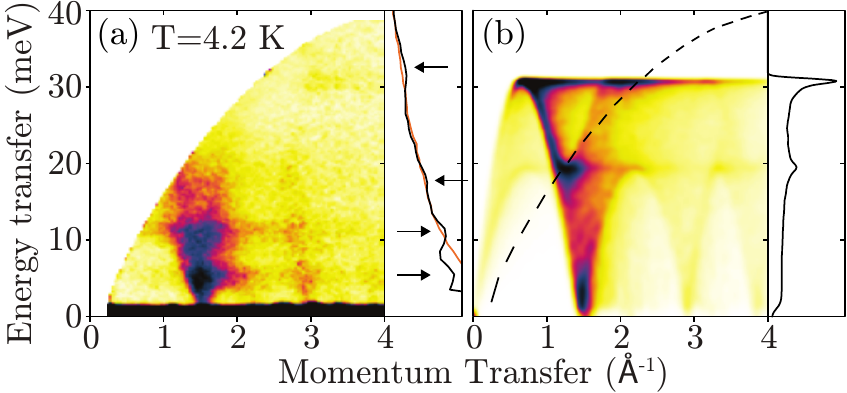}
		\caption{(color online) (a) Energy and wavevector dependence of ARCS spectra for $E_\text{i}  = 40$ meV. Right side shows an energy cut for $1.5 \leq Q \leq 3.3$ \AA\ from the $E_\text{i}=60$ meV data measured at 4.2 K (black line) and 55 K (orange line). Arrows indicate the van Hove singularities at 4.2 K. (b) Powder average of calculated spectra for a spatially isotropic triangular lattice with interplane coupling, dashed line indicates the border of the measured region. The right side shows the equivalent energy cut. Intensities are in arbitrary units.}
		\label{fig:arcs}
	\end{figure}

	Powder inelastic neutron scattering was measured on the ARCS \cite{arcs} chopper spectrometer at SNS, ORNL, USA. Data was collected with incident neutron energies of 40 meV and 60 meV (resolution of 1.64 meV and 2.35 meV at the elastic line respectively) at 4.2 K and 55 K.	To map out the excitations along all reciprocal space directions, a single crystal experiment was performed on the Merlin \cite{Bewley06}  chopper spectrometer at the ISIS facility, RAL, UK. The sample was measured at 5 K using an incident neutron energy of 50 meV. The crystal was rotated over 120$^\circ$ in 1$^\circ$ steps about the vertical $a$ axis and data was collected for 1 hour per angle. The data was transformed into units of energy and wavevector transfer, combined and integrated over $h$ using the Horace software \cite{Horace}.	Single crystal experiments were also performed on the Puma thermal triple axis spectrometer at FRM2, Germany. The final neutron wavelength was fixed to 2.662 \AA$^{-1}$, giving an energy resolution of 0.75 meV. The magnetic excitations were measured at 3.5 K, while several scans were repeated at 220 K to separate phonon and magnon scattering. To determine the strength of the weak interplane coupling and check for an anisotropy gap, scans were performed on the cold neutron high resolution V2 triple axis spectrometer at HZB, Germany. A fixed final neutron wavelength of  1.20 \AA$^{-1}$ gave an energy resolution of 0.093 meV.

%%%%%%%%%%%%%%%%%%%%%%%%%%%%%%%%%%%%%%%%%%%%%%%%%%%%%%%%%%%%%%%%%%%%%%%%%%%%%%%%%%%%%%%%%%%%%%%%%%%%%%%%%%%%
% Analysis
%%%%%%%%%%%%%%%%%%%%%%%%%%%%%%%%%%%%%%%%%%%%%%%%%%%%%%%%%%%%%%%%%%%%%%%%%%%%%%%%%%%%%%%%%%%%%%%%%%%%%%%%%%%%

	The powder spectra shows strong inelastic intensity centered around Q$\approx$1.54 \AA$^{-1}$ corresponding to the (1,4/3,0) magnetic Bragg-peak (1.55 \AA$^{-1}$), see Fig.\ \ref{fig:arcs}(a), where the characteristic V-shaped dispersion is visible. Van Hove singularities are observed at 5, 11, 17 and 33 meV indicating energies where the dispersion is flat, however no sharp upper edge of the dispersion is observed as would be naively expected in the classical case. The data can be compared to a spin wave simulation, assuming spatially isotropic nearest-neighbor interactions $J_\text{nn}=$6.48 meV (from the high temperature susceptibility) and interplane coupling $J_\text{int}=$0.005 meV. The simulated spectra in Fig.\ \ref{fig:arcs}(b) shows van Hove singularities at 19 meV and at the top of the magnon dispersion at 32 meV. These singularities are missing in the measured spectra and the magnon intensity decreases smoothly to background with increasing energy transfer. It is clear that despite its 120$^\circ$ magnetic structure the excitations of $\alpha$-CaCr$_2$O$_4$ cannot be approximated by an isotropic triangular antiferromagnet and the spatial anisotropy of the interactions must be considered.
	\begin{figure}[!htb]
		\centering
		\includegraphics[width= 86 mm]{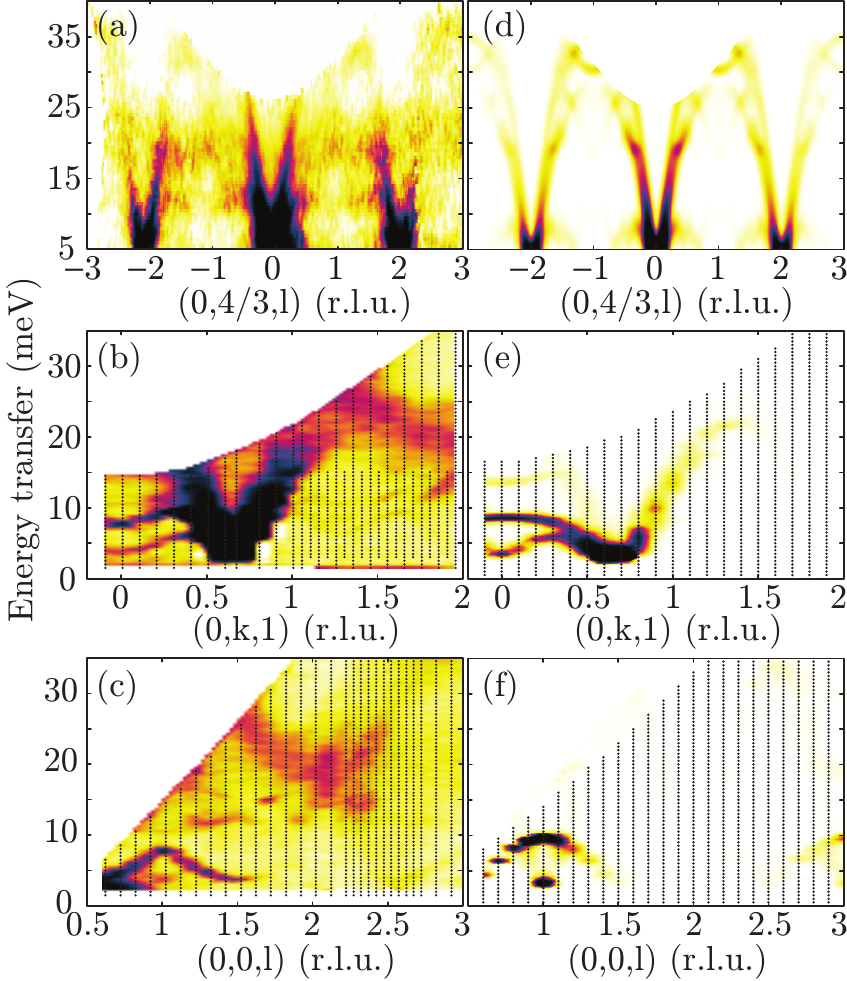}
		\caption{(color online) (a) Spectra measured on Merlin, (b-c) spectra measured on Puma, black dots denote measurement points. (d-f) Calculated spin-wave spectra using the best fit parameters (Tab.\ \ref{tab:Jval}). The instrumental resolution is convolved with the calculated spectra, using the resolution ellipsoid obtained from Rescal \cite{Rescal} (Popovici's method \cite{popovici75}). To reproduce all the observed modes, the dispersion of the three crystallographic twins were combined, weighted by their experimentally determined volume fractions. Intensities are in arbitrary units, , ``r.l.u.'' stands for reciprocal lattice units.}
		\label{fig:puma}
	\end{figure}

	The full complexity of the spectra is revealed by the single crystal data. Figure \ref{fig:puma} (a) shows the (0,4/3,$l$) dispersion from Merlin. The scattering intensity is strong around the magnetic zone centers (0,4/3,$\pm$2) and (0,4/3,0) as expected. Many spin-wave branches disperse from these points, the steepest of which reach a zone boundary energy of 35 meV. Details of the lower energy branches are revealed by Puma. Along (0,k,1) (Fig.\ \ref{fig:puma}(b)), the strong scattering around the zone center is again visible at (0,2/3,1), the V-shaped cone is  typical of the magnon dispersion of the spatially isotropic triangular lattice. However it is broader than the resolution probably due to multiple spin-wave branches arising from having eight magnetic ions in the unit cell and three crystallographic twins. The gap at (0,2/3,1) is due to the interplane coupling. On both sides of the cone, low energy modes ($E \leq 10$) are visible with local minima at (0,0,1) and (0,4/3,1). The scattering in the region $1.3 < k < 2$ and $15 < E < 25$ is from phonons as confirmed by comparing to the spectra at high temperatures ($T$=220 K). The low energy modes are also visible along (0,0,$l$), (Fig.\ \ref{fig:puma} (c)). They are especially strong around $l\approx1$ and have a distinctive shape. The broad modes above 15 meV are again phonons. 

	In order to probe the interplane coupling, energy scans at the zone center were repeated at different values of out-of-plane wavevector. The scan at the (1,4/3,0) magnetic Bragg peak shows no gap revealing that anisotropy is negligible. However the magnetic excitations at (2,4/3,0) have a gap of 3.5 meV revealing a small but significant antiferromagnetic coupling between adjacent planes. 

%%%%%%%%%%%%%%%%%%%%%%%%%%%%%%%%%%%%%%%%%%%%%%%%%%%%%%%%%%%%%%%%%%%%%%%%%%%%%%%%%%%%%%%%%%%%%%%%%%%%%%%%%%%%
% Modelling
%%%%%%%%%%%%%%%%%%%%%%%%%%%%%%%%%%%%%%%%%%%%%%%%%%%%%%%%%%%%%%%%%%%%%%%%%%%%%%%%%%%%%%%%%%%%%%%%%%%%%%%%%%%%

	A spin-wave model was developed to explain the observed features in the excitation spectra of $\alpha$-CaCr$_2$O$_4$. Based on the crystal structure and magnetization data, the magnetic properties should be well accounted for by a Heisenberg model. Four inequivalent nearest neighbor and four inequivalent next-nearest-neighbor interactions in the triangular plane are allowed by symmetry. The interplane exchange interactions are assumed to be all equal. We consider the spin Hamiltonian to be composed of interplane, nearest-neighbor and next-nearest-neighbor terms. Since the system is strongly frustrated, the magnetic ordering wavevector and spin directions are sensitive to the sizes of the interactions, for details see Ref. \onlinecite{Toth2011}. The observed value of the ordering wavevector $k_\text{m}$=(0,1/3,0) gives the following restriction on the parameters:
	\begin{align}
		\label{eq:d3}
		\Delta_3 = J_\text{zz1}+J_\text{zz2}-J_\text{ch1}-J_\text{ch2} = 0,
	\end{align}
where the zz and ch indices denote the zig-zag and chain nearest-neighbor in-plane interactions respectively. The next-nearest-neighbor interactions ($J_\text{nnn1}$-$J_\text{nnn4}$) do not influence the ordering wavevector. The above constraint is applied in the rest of this paper. We define a set of parameters linearly independent of $\Delta_3$:
	\begin{align}
		&J_\text{nn}=\frac{1}{4}(J_\text{ch1}+J_\text{ch2}+J_\text{zz1}+J_\text{zz2}),        \nonumber\\
		\Delta_1 =\frac{1}{2J_\text{nn}}& (J_\text{zz1}-J_\text{zz2}),\;\;\; \Delta_2 =\frac{1}{2J_\text{nn}} (J_\text{ch1}-J_\text{ch2}).
	\end{align}
The magnetic structure as a function of the new parameters is determined by minimizing the energy, assuming a single pair of modulation wave vectors $k_\text{m}$ and $-k_\text{m}$ \cite{Nagamiya67}.

	Figure \ref{fig:struct}(c-e) shows the phase diagram as a function of $\Delta_1$, $\Delta_2$ and average next-nearest-neighbor interaction $J_\text{nnn}$. The 120$^\circ$ structure characterized by $k_\text{m}$=(0,1/3,0) is stable over a large range of exchange constants. In the shaded region the magnetic ground state has a multi-$k$ structure, which is beyond our current model. A Monte Carlo method applied to an extended unit cell would be necessary to determine the structure here. Beyond the shaded region the magnetic structure can again be described by a single wavevector but with a new value e.g.\ $k_\text{m}$=(0,0,1/2) represented by the circle in Fig.\ \ref{fig:struct}(d) and illustrated in Fig.\ \ref{fig:struct}(b)
	\begin{ruledtabular}
	\begin{table}[!htb]
		\centering
		\caption{Values of the fitted exchange parameters.}
		\label{tab:Jval}
		\begin{tabular}{l|l|l|l|l|l}
			Name               & $J_\text{zz1}$  & $J_\text{zz2}$  & $J_\text{ch1}$  & $J_\text{ch2}$  & $J_\text{int}$\\\hline
			Value (meV)        & 5.8(6)          & 11.8(14)        & 9.1(10)         & 8.6(15)         & 0.027(1)      \\\hline
			Cr--Cr dist. (\AA) & 2.939           & 2.889           & 2.911           & 2.907           & 5.529         \\\hline
			Name               & $J_\text{nnn1}$ & $J_\text{nnn2}$ & $J_\text{nnn3}$ & $J_\text{nnn4}$ &               \\\hline
			Value (meV)        & 0.11            & 0.31            & 0.96            & 0.90            &               \\
		\end{tabular}
	\end{table}
	\end{ruledtabular}

Linear spin-wave theory was used to calculate the excitation spectra and neutron scattering cross section of the proposed Hamiltonian \cite{Chernyshev2009,White1965,coldeathesis}. LSWT fails if the magnetic ground state structure is incorrect for a given set of exchange constants, and since a single pair of ordering wavevectors is assumed in the spin-wave calculation, the borders of the multi-$k$ phase can be determined by the values of the exchange interactions where the spin-wave energies become imaginary, (shaded regions in Fig.\ \ref{fig:struct}(c-e). Using the parameter constraint $\Delta_3 = 0$, we fitted the parameters $J_\text{nn}$, $\Delta_1$, $\Delta_2$, $J_\text{int}$, $J_\text{nnn1}$, $J_\text{nnn2}$, $J_\text{nnn3}$, $J_\text{nnn4}$. The calculated modes with strongest intensity were fitted to the dispersion curves extracted from the data.	The fit favored a set of exchange interactions close to the boundary between the $k_\text{m}$=(0,1/3,0) and multi-$k$ phase. In fact the fit was unstable, driving the exchange parameters out of the 120$^\circ$ region, and it had to be constrained to stay within this phase to ensure consistency with the diffraction results. The best fits were obtained with the following values: $J_\text{nn} = 8.8(8)$ meV, $\Delta_1=\pm0.34(4)$, $\Delta_2=0.03(6)$ and the average next-nearest-neighbor strength $J_\text{nnn}=0.69$ meV. Since the four inequivalent next-nearest-neighbor interactions are strongly correlated, no unique set of values were found. A good solution is shown in Tab.\ \ref{tab:Jval} and the corresponding fitted dispersion relations are shown alongside the data in Fig.\ \ref{fig:puma}

The values of $\Delta_1$ and $\Delta_2$ correlate well with the Cr$^{3+}$--Cr$^{3+}$ distances determined by neutron powder diffraction. The larger $\Delta_1$ value arises from the difference between the two zig-zag interaction strengths, where the difference between their distances is 0.050 \AA. In contrast, the difference between the two chain distances is only 0.004 \AA, thus explaining the very small difference between their interactions ($\Delta_2$). The sign of $\Delta_1$ is undetermined, because it does not affect the spin-wave spectra, however to make the shorter coupling stronger $\Delta_1$ has been set negative. These values for the first neighbor exchange constants agree well with the Cr$^{3+}$--Cr$^{3+}$ interaction strength as a function of distance as found for other chromium oxides delafossites with direct exchange interactions \cite{Poienar2010, Frontzek2011, Kan2009, Hsieh2008a, Moreno2004, Delmas1978, Meisenheimer1961, Takatsu2009}.

The results of the LSWT fit place $\alpha$-CaCr$_2$O$_4$ at the edge of the 120$^\circ$ phase close to the multi-$k$ phase, as shown by the circle in Fig.\ \ref{fig:struct}(d). The proximity of the phase boundary is revealed by the presence of low energy modes with roton-like minima at wavevectors that do not correspond to the $k_\text{m} = (0,1/3,0)$ phase but rather act as soft modes of the nearby structure. LSWT is able to reproduce the spectra but there are small discrepancies (Fig.\ \ref{fig:puma}). It has been shown theoretically that for the isotropic triangular AF the non-collinearity of the spin moments gives rise to substantial downwards renormalization and broadening compared to LSWT and while these effects are most pronounced for spin-1/2 \cite{Starykh2006,Zheng2006,Chernyshev2009}, recent calculations suggest that they may be $\sim$10 \% for spin-3/2 \cite{Chernyshevpc}. Quantum effects are also enhanced for low energy modes close to a phase boundary. Soft modes are predicted for the collinear phase of the isotropic triangular lattice with next-nearest-neighbor interactions \cite{Wheeler2009a,Chubukov1992} corresponding to several possible phases which are classically degenerate. Via the mechanism of order by disorder quantum fluctuations (QF) select a striped collinear state by keeping the relevant soft mode gapless while increasing the gap of the other modes. It was also shown, that QF can shift the magnetic phase boundary of the isotropic triangular AF with next-nearest-neighbor interactions in external magnetic field \cite{Fishman2011}. 

Since QF have a significant effect on the isotropic triangular AF close to phase boundaries, it is possible that they also modify the spectra of $\alpha$-CaCr$_2$O$_4$. These modifications might be responsible for stabilizing the 120$^\circ$ phase via the mechanism of order by disorder even though the LSWT fit to the dispersion suggests that $\alpha$-CaCr$_2$O$_4$ might lie outside this phase. It should be noted that the isostructural compound $\alpha$-SrCr$_2$O$_4$ also has a 120$^\circ$ magnetic structure, here the crystal structure suggests that the difference between chain interactions is greater than in $\alpha$-CaCr$_2$O$_4$ while the difference between the zig-zags in smaller \cite{Dutton11}. The excitation spectra of this compound has not been measured. Further theoretical study taking quantum fluctuations into account is necessary to determine the phase boundary and excitation spectra.
	
	In conclusion, despite its highly symmetric 120$^\circ$ magnetic structure, the magnetic excitation spectrum of $\alpha$-CaCr$_2$O$_4$ is highly complex. The unique pattern of nearest neighbor exchange interactions as well as the substantial next-nearest-neighbor interactions place it close to the phase boundary of the 120$^\circ$ structure as is clearly revealed by the presence of low energy modes acting as soft-modes of the neighboring structure. Indeed the LSWT fit tends to drive the system towards the nearby multi-$k$ phase in contradiction to the observed ordering wavevector of $k_\text{m}$=(0,1/3,0) and quantum fluctuations may be necessary to stabilize the system at a well-defined point inside the 120$^\circ$ phase.

%%%%%%%%%%%%%%%%%%%%%%%%%%%%%%%%%%%%%%%%%%%%%%%%%%%%%%%%%%%%%%%%%%%%%%%%%%%%%%%%%%%%%%%%%%%%%%%%%%%%%%%%%%%%
% Acknowledgement
%%%%%%%%%%%%%%%%%%%%%%%%%%%%%%%%%%%%%%%%%%%%%%%%%%%%%%%%%%%%%%%%%%%%%%%%%%%%%%%%%%%%%%%%%%%%%%%%%%%%%%%%%%%%

	We thank A. J. Williams for the powder sample, O. Pieper and N. Shannon for fruitful discussions. The Research at Oak Ridge National Laboratory's Spallation Neutron Source was sponsored by the Scientific User Facilities Division, Office of Basic Energy Sciences, U. S. Department of Energy.

\bibliography{bibliography}

\begin{thebibliography}{32}%
\makeatletter
\providecommand \@ifxundefined [1]{%
 \@ifx{#1\undefined}
}%
\providecommand \@ifnum [1]{%
 \ifnum #1\expandafter \@firstoftwo
 \else \expandafter \@secondoftwo
 \fi
}%
\providecommand \@ifx [1]{%
 \ifx #1\expandafter \@firstoftwo
 \else \expandafter \@secondoftwo
 \fi
}%
\providecommand \natexlab [1]{#1}%
\providecommand \enquote  [1]{``#1''}%
\providecommand \bibnamefont  [1]{#1}%
\providecommand \bibfnamefont [1]{#1}%
\providecommand \citenamefont [1]{#1}%
\providecommand \href@noop [0]{\@secondoftwo}%
\providecommand \href [0]{\begingroup \@sanitize@url \@href}%
\providecommand \@href[1]{\@@startlink{#1}\@@href}%
\providecommand \@@href[1]{\endgroup#1\@@endlink}%
\providecommand \@sanitize@url [0]{\catcode `\\12\catcode `\$12\catcode
  `\&12\catcode `\#12\catcode `\^12\catcode `\_12\catcode `\%12\relax}%
\providecommand \@@startlink[1]{}%
\providecommand \@@endlink[0]{}%
\providecommand \url  [0]{\begingroup\@sanitize@url \@url }%
\providecommand \@url [1]{\endgroup\@href {#1}{\urlprefix }}%
\providecommand \urlprefix  [0]{URL }%
\providecommand \Eprint [0]{\href }%
\providecommand \doibase [0]{http://dx.doi.org/}%
\providecommand \selectlanguage [0]{\@gobble}%
\providecommand \bibinfo  [0]{\@secondoftwo}%
\providecommand \bibfield  [0]{\@secondoftwo}%
\providecommand \translation [1]{[#1]}%
\providecommand \BibitemOpen [0]{}%
\providecommand \bibitemStop [0]{}%
\providecommand \bibitemNoStop [0]{.\EOS\space}%
\providecommand \EOS [0]{\spacefactor3000\relax}%
\providecommand \BibitemShut  [1]{\csname bibitem#1\endcsname}%
\let\auto@bib@innerbib\@empty
%</preamble>
\bibitem [{\citenamefont {Ramirez}(2001)}]{ramirez2001}%
  \BibitemOpen
  \bibfield  {author} {\bibinfo {author} {\bibfnamefont {A.~P.}\ \bibnamefont
  {Ramirez}},\ }\href@noop {} {\emph {\bibinfo {title} {Handbook of magnetic
  materials}}},\ edited by\ \bibinfo {editor} {\bibfnamefont {K.~J.~H.}\
  \bibnamefont {Buschow}},\ Vol.~\bibinfo {volume} {13}\ (\bibinfo  {publisher}
  {Elsevier Science},\ \bibinfo {address} {Amsterdam},\ \bibinfo {year}
  {2001})\ p.\ \bibinfo {pages} {423}\BibitemShut {NoStop}%
\bibitem [{\citenamefont {Moessner}\ and\ \citenamefont
  {Sondhi}(2001)}]{Moessner2001}%
  \BibitemOpen
  \bibfield  {author} {\bibinfo {author} {\bibfnamefont {R.}~\bibnamefont
  {Moessner}}\ and\ \bibinfo {author} {\bibfnamefont {S.~L.}\ \bibnamefont
  {Sondhi}},\ }\href {\doibase 10.1103/PhysRevLett.86.1881} {\bibfield
  {journal} {\bibinfo  {journal} {Phys. Rev. Lett.}\ }\textbf {\bibinfo
  {volume} {86}},\ \bibinfo {pages} {1881} (\bibinfo {year}
  {2001})}\BibitemShut {NoStop}%
\bibitem [{\citenamefont {Tsunetsugu}\ and\ \citenamefont
  {Arikawa}(2006)}]{Tsunetsugu2006}%
  \BibitemOpen
  \bibfield  {author} {\bibinfo {author} {\bibfnamefont {H.}~\bibnamefont
  {Tsunetsugu}}\ and\ \bibinfo {author} {\bibfnamefont {M.}~\bibnamefont
  {Arikawa}},\ }\href {\doibase 10.1143/JPSJ.75.083701} {\bibfield  {journal}
  {\bibinfo  {journal} {J. Phys. Soc. Jpn.}\ }\textbf {\bibinfo {volume}
  {75}},\ \bibinfo {pages} {083701} (\bibinfo {year} {2006})}\BibitemShut
  {NoStop}%
\bibitem [{\citenamefont {Starykh}\ \emph {et~al.}(2006)\citenamefont
  {Starykh}, \citenamefont {Chubukov},\ and\ \citenamefont
  {Abanov}}]{Starykh2006}%
  \BibitemOpen
  \bibfield  {author} {\bibinfo {author} {\bibfnamefont {O.~A.}\ \bibnamefont
  {Starykh}}, \bibinfo {author} {\bibfnamefont {A.~V.}\ \bibnamefont
  {Chubukov}}, \ and\ \bibinfo {author} {\bibfnamefont {A.~G.}\ \bibnamefont
  {Abanov}},\ }\href {\doibase 10.1103/PhysRevB.74.180403} {\bibfield
  {journal} {\bibinfo  {journal} {Phys. Rev. B}\ }\textbf {\bibinfo {volume}
  {74}},\ \bibinfo {pages} {180403} (\bibinfo {year} {2006})}\BibitemShut
  {NoStop}%
\bibitem [{\citenamefont {Zheng}\ \emph {et~al.}(2006)\citenamefont {Zheng},
  \citenamefont {Fj\ae{}restad}, \citenamefont {Singh}, \citenamefont
  {McKenzie},\ and\ \citenamefont {Coldea}}]{Zheng2006}%
  \BibitemOpen
  \bibfield  {author} {\bibinfo {author} {\bibfnamefont {W.}~\bibnamefont
  {Zheng}}, \bibinfo {author} {\bibfnamefont {J.~O.}\ \bibnamefont
  {Fj\ae{}restad}}, \bibinfo {author} {\bibfnamefont {R.~R.~P.}\ \bibnamefont
  {Singh}}, \bibinfo {author} {\bibfnamefont {R.~H.}\ \bibnamefont {McKenzie}},
  \ and\ \bibinfo {author} {\bibfnamefont {R.}~\bibnamefont {Coldea}},\ }\href
  {\doibase 10.1103/PhysRevLett.96.057201} {\bibfield  {journal} {\bibinfo
  {journal} {Phys. Rev. Lett.}\ }\textbf {\bibinfo {volume} {96}},\ \bibinfo
  {pages} {057201} (\bibinfo {year} {2006})}\BibitemShut {NoStop}%
\bibitem [{\citenamefont {Chernyshev}\ and\ \citenamefont
  {Zhitomirsky}(2009)}]{Chernyshev2009}%
  \BibitemOpen
  \bibfield  {author} {\bibinfo {author} {\bibfnamefont {A.~L.}\ \bibnamefont
  {Chernyshev}}\ and\ \bibinfo {author} {\bibfnamefont {M.~E.}\ \bibnamefont
  {Zhitomirsky}},\ }\href {\doibase 10.1103/PhysRevB.79.144416} {\bibfield
  {journal} {\bibinfo  {journal} {Phys. Rev. B}\ }\textbf {\bibinfo {volume}
  {79}},\ \bibinfo {eid} {144416} (\bibinfo {year} {2009})}\BibitemShut
  {NoStop}%
\bibitem [{\citenamefont {Wheeler}\ \emph {et~al.}(2009)\citenamefont
  {Wheeler}, \citenamefont {Coldea}, \citenamefont {Wawrzy\'{n}ska},
  \citenamefont {S\"{o}rgel}, \citenamefont {Jansen}, \citenamefont {Koza},
  \citenamefont {Taylor}, \citenamefont {Adroguer},\ and\ \citenamefont
  {Shannon}}]{Wheeler2009a}%
  \BibitemOpen
  \bibfield  {author} {\bibinfo {author} {\bibfnamefont {E.~M.}\ \bibnamefont
  {Wheeler}}, \bibinfo {author} {\bibfnamefont {R.}~\bibnamefont {Coldea}},
  \bibinfo {author} {\bibfnamefont {E.}~\bibnamefont {Wawrzy\'{n}ska}},
  \bibinfo {author} {\bibfnamefont {T.}~\bibnamefont {S\"{o}rgel}}, \bibinfo
  {author} {\bibfnamefont {M.}~\bibnamefont {Jansen}}, \bibinfo {author}
  {\bibfnamefont {M.~M.}\ \bibnamefont {Koza}}, \bibinfo {author}
  {\bibfnamefont {J.}~\bibnamefont {Taylor}}, \bibinfo {author} {\bibfnamefont
  {P.}~\bibnamefont {Adroguer}}, \ and\ \bibinfo {author} {\bibfnamefont
  {N.}~\bibnamefont {Shannon}},\ }\href {\doibase 10.1103/PhysRevB.79.104421}
  {\bibfield  {journal} {\bibinfo  {journal} {Phys. Rev. B}\ }\textbf {\bibinfo
  {volume} {79}},\ \bibinfo {pages} {104421} (\bibinfo {year}
  {2009})}\BibitemShut {NoStop}%
\bibitem [{\citenamefont {Chubukov}\ and\ \citenamefont
  {Jolicoeur}(1992)}]{Chubukov1992}%
  \BibitemOpen
  \bibfield  {author} {\bibinfo {author} {\bibfnamefont {A.~V.}\ \bibnamefont
  {Chubukov}}\ and\ \bibinfo {author} {\bibfnamefont {T.}~\bibnamefont
  {Jolicoeur}},\ }\href {\doibase 10.1103/PhysRevB.46.11137} {\bibfield
  {journal} {\bibinfo  {journal} {Phys. Rev. B}\ }\textbf {\bibinfo {volume}
  {46}},\ \bibinfo {pages} {11137} (\bibinfo {year} {1992})}\BibitemShut
  {NoStop}%
\bibitem [{\citenamefont {Coldea}\ \emph {et~al.}(1996)\citenamefont {Coldea},
  \citenamefont {Tennant}, \citenamefont {Cowley}, \citenamefont {McMorrow},
  \citenamefont {Dorner},\ and\ \citenamefont {Tylczynski}}]{Coldea1996}%
  \BibitemOpen
  \bibfield  {author} {\bibinfo {author} {\bibfnamefont {R.}~\bibnamefont
  {Coldea}}, \bibinfo {author} {\bibfnamefont {D.~A.}\ \bibnamefont {Tennant}},
  \bibinfo {author} {\bibfnamefont {R.~A.}\ \bibnamefont {Cowley}}, \bibinfo
  {author} {\bibfnamefont {D.~F.}\ \bibnamefont {McMorrow}}, \bibinfo {author}
  {\bibfnamefont {B.}~\bibnamefont {Dorner}}, \ and\ \bibinfo {author}
  {\bibfnamefont {Z.}~\bibnamefont {Tylczynski}},\ }\href {\doibase
  10.1088/0953-8984/8/40/012} {\bibfield  {journal} {\bibinfo  {journal} {J.
  Phys. Condens. Matter}\ }\textbf {\bibinfo {volume} {8}},\ \bibinfo {pages}
  {7473} (\bibinfo {year} {1996})}\BibitemShut {NoStop}%
\bibitem [{\citenamefont {Poienar}\ \emph {et~al.}(2010)\citenamefont
  {Poienar}, \citenamefont {Damay}, \citenamefont {Martin}, \citenamefont
  {Robert},\ and\ \citenamefont {Petit}}]{Poienar2010}%
  \BibitemOpen
  \bibfield  {author} {\bibinfo {author} {\bibfnamefont {M.}~\bibnamefont
  {Poienar}}, \bibinfo {author} {\bibfnamefont {F.}~\bibnamefont {Damay}},
  \bibinfo {author} {\bibfnamefont {C.}~\bibnamefont {Martin}}, \bibinfo
  {author} {\bibfnamefont {J.}~\bibnamefont {Robert}}, \ and\ \bibinfo {author}
  {\bibfnamefont {S.}~\bibnamefont {Petit}},\ }\href
  {http://link.aps.org/doi/10.1103/PhysRevB.81.104411} {\bibfield  {journal}
  {\bibinfo  {journal} {Phys. Rev. B}\ }\textbf {\bibinfo {volume} {81}},\
  \bibinfo {pages} {104411} (\bibinfo {year} {2010})}\BibitemShut {NoStop}%
\bibitem [{\citenamefont {Pausch}\ \emph {et~al.}(1974)\citenamefont {Pausch},
  \citenamefont {Mueller-Buschbaum},\ and\ \citenamefont
  {Pausch}}]{Pausch1974}%
  \BibitemOpen
  \bibfield  {author} {\bibinfo {author} {\bibfnamefont {H.}~\bibnamefont
  {Pausch}}, \bibinfo {author} {\bibfnamefont {H.}~\bibnamefont
  {Mueller-Buschbaum}}, \ and\ \bibinfo {author} {\bibfnamefont {V.~H.}\
  \bibnamefont {Pausch}},\ }\href {\doibase 10.1002/zaac.19744050113}
  {\bibfield  {journal} {\bibinfo  {journal} {Z. Anorg. Allg. Chem.}\ }\textbf
  {\bibinfo {volume} {405}},\ \bibinfo {pages} {113} (\bibinfo {year}
  {1974})}\BibitemShut {NoStop}%
\bibitem [{\citenamefont {Dutton}\ \emph {et~al.}(2011)\citenamefont {Dutton},
  \citenamefont {Climent-Pascual}, \citenamefont {Stephens}, \citenamefont
  {Hodges}, \citenamefont {Huq}, \citenamefont {Broholm},\ and\ \citenamefont
  {Cava}}]{Dutton11}%
  \BibitemOpen
  \bibfield  {author} {\bibinfo {author} {\bibfnamefont {S.~E.}\ \bibnamefont
  {Dutton}}, \bibinfo {author} {\bibfnamefont {E.}~\bibnamefont
  {Climent-Pascual}}, \bibinfo {author} {\bibfnamefont {P.~W.}\ \bibnamefont
  {Stephens}}, \bibinfo {author} {\bibfnamefont {J.~P.}\ \bibnamefont
  {Hodges}}, \bibinfo {author} {\bibfnamefont {A.}~\bibnamefont {Huq}},
  \bibinfo {author} {\bibfnamefont {C.~L.}\ \bibnamefont {Broholm}}, \ and\
  \bibinfo {author} {\bibfnamefont {R.~J.}\ \bibnamefont {Cava}},\ }\href
  {http://stacks.iop.org/0953-8984/23/i=24/a=246005} {\bibfield  {journal}
  {\bibinfo  {journal} {J. Phys. Condens. Matter}\ }\textbf {\bibinfo {volume}
  {23}},\ \bibinfo {pages} {246005} (\bibinfo {year} {2011})}\BibitemShut
  {NoStop}%
\bibitem [{\citenamefont {Frontzek}\ \emph {et~al.}(2011)\citenamefont
  {Frontzek}, \citenamefont {Haraldsen}, \citenamefont {Podlesnyak},
  \citenamefont {Matsuda}, \citenamefont {Christianson}, \citenamefont
  {Fishman}, \citenamefont {Sefat}, \citenamefont {Qiu}, \citenamefont
  {Copley}, \citenamefont {Barilo}, \citenamefont {Shiryaev},\ and\
  \citenamefont {Ehlers}}]{Frontzek2011}%
  \BibitemOpen
  \bibfield  {author} {\bibinfo {author} {\bibfnamefont {M.}~\bibnamefont
  {Frontzek}}, \bibinfo {author} {\bibfnamefont {J.~T.}\ \bibnamefont
  {Haraldsen}}, \bibinfo {author} {\bibfnamefont {A.}~\bibnamefont
  {Podlesnyak}}, \bibinfo {author} {\bibfnamefont {M.}~\bibnamefont {Matsuda}},
  \bibinfo {author} {\bibfnamefont {A.~D.}\ \bibnamefont {Christianson}},
  \bibinfo {author} {\bibfnamefont {R.~S.}\ \bibnamefont {Fishman}}, \bibinfo
  {author} {\bibfnamefont {A.~S.}\ \bibnamefont {Sefat}}, \bibinfo {author}
  {\bibfnamefont {Y.}~\bibnamefont {Qiu}}, \bibinfo {author} {\bibfnamefont
  {J.~R.~D.}\ \bibnamefont {Copley}}, \bibinfo {author} {\bibfnamefont
  {S.}~\bibnamefont {Barilo}}, \bibinfo {author} {\bibfnamefont {S.~V.}\
  \bibnamefont {Shiryaev}}, \ and\ \bibinfo {author} {\bibfnamefont
  {G.}~\bibnamefont {Ehlers}},\ }\href {\doibase 10.1103/PhysRevB.84.094448}
  {\bibfield  {journal} {\bibinfo  {journal} {Phys. Rev. B}\ }\textbf {\bibinfo
  {volume} {84}},\ \bibinfo {pages} {094448} (\bibinfo {year}
  {2011})}\BibitemShut {NoStop}%
\bibitem [{\citenamefont {Kan}\ \emph {et~al.}(2009)\citenamefont {Kan},
  \citenamefont {Xiang}, \citenamefont {Zhang}, \citenamefont {Lee},\ and\
  \citenamefont {Whangbo}}]{Kan2009}%
  \BibitemOpen
  \bibfield  {author} {\bibinfo {author} {\bibfnamefont {E.~J.}\ \bibnamefont
  {Kan}}, \bibinfo {author} {\bibfnamefont {H.~J.}\ \bibnamefont {Xiang}},
  \bibinfo {author} {\bibfnamefont {Y.}~\bibnamefont {Zhang}}, \bibinfo
  {author} {\bibfnamefont {C.}~\bibnamefont {Lee}}, \ and\ \bibinfo {author}
  {\bibfnamefont {M.-H.}\ \bibnamefont {Whangbo}},\ }\href {\doibase
  10.1103/PhysRevB.80.104417} {\bibfield  {journal} {\bibinfo  {journal} {Phys.
  Rev. B}\ }\textbf {\bibinfo {volume} {80}},\ \bibinfo {pages} {104417}
  (\bibinfo {year} {2009})}\BibitemShut {NoStop}%
\bibitem [{\citenamefont {Hsieh}\ \emph {et~al.}(2008)\citenamefont {Hsieh},
  \citenamefont {Qian}, \citenamefont {Berger}, \citenamefont {Cava},
  \citenamefont {Lynn}, \citenamefont {Huang},\ and\ \citenamefont
  {Hasan}}]{Hsieh2008a}%
  \BibitemOpen
  \bibfield  {author} {\bibinfo {author} {\bibfnamefont {D.}~\bibnamefont
  {Hsieh}}, \bibinfo {author} {\bibfnamefont {D.}~\bibnamefont {Qian}},
  \bibinfo {author} {\bibfnamefont {R.}~\bibnamefont {Berger}}, \bibinfo
  {author} {\bibfnamefont {R.}~\bibnamefont {Cava}}, \bibinfo {author}
  {\bibfnamefont {J.}~\bibnamefont {Lynn}}, \bibinfo {author} {\bibfnamefont
  {Q.}~\bibnamefont {Huang}}, \ and\ \bibinfo {author} {\bibfnamefont
  {M.}~\bibnamefont {Hasan}},\ }\href {\doibase 10.1016/j.jpcs.2008.06.124}
  {\bibfield  {journal} {\bibinfo  {journal} {J. Phys. Chem. Solids}\ }\textbf
  {\bibinfo {volume} {69}},\ \bibinfo {pages} {3174 } (\bibinfo {year}
  {2008})}\BibitemShut {NoStop}%
\bibitem [{\citenamefont {Moreno}\ \emph {et~al.}(2004)\citenamefont {Moreno},
  \citenamefont {Israel}, \citenamefont {Pagliuso}, \citenamefont
  {Garcia-Adeva}, \citenamefont {Rettori}, \citenamefont {Sarrao},
  \citenamefont {Thompson},\ and\ \citenamefont {Oseroff}}]{Moreno2004}%
  \BibitemOpen
  \bibfield  {author} {\bibinfo {author} {\bibfnamefont {N.~O.}\ \bibnamefont
  {Moreno}}, \bibinfo {author} {\bibfnamefont {C.}~\bibnamefont {Israel}},
  \bibinfo {author} {\bibfnamefont {P.~G.}\ \bibnamefont {Pagliuso}}, \bibinfo
  {author} {\bibfnamefont {A.~J.}\ \bibnamefont {Garcia-Adeva}}, \bibinfo
  {author} {\bibfnamefont {C.}~\bibnamefont {Rettori}}, \bibinfo {author}
  {\bibfnamefont {J.~L.}\ \bibnamefont {Sarrao}}, \bibinfo {author}
  {\bibfnamefont {J.~D.}\ \bibnamefont {Thompson}}, \ and\ \bibinfo {author}
  {\bibfnamefont {S.~B.}\ \bibnamefont {Oseroff}},\ }\href {\doibase
  10.1016/j.jmmm.2003.12.122} {\bibfield  {journal} {\bibinfo  {journal} {J.
  Magn. Magn. Mater.}\ }\textbf {\bibinfo {volume} {272-276}},\ \bibinfo
  {pages} {E1023 } (\bibinfo {year} {2004})}\BibitemShut {NoStop}%
\bibitem [{\citenamefont {Delmas}\ \emph {et~al.}(1978)\citenamefont {Delmas},
  \citenamefont {Flem}, \citenamefont {Fouassier},\ and\ \citenamefont
  {Hagenmuller}}]{Delmas1978}%
  \BibitemOpen
  \bibfield  {author} {\bibinfo {author} {\bibfnamefont {C.}~\bibnamefont
  {Delmas}}, \bibinfo {author} {\bibfnamefont {G.~L.}\ \bibnamefont {Flem}},
  \bibinfo {author} {\bibfnamefont {C.}~\bibnamefont {Fouassier}}, \ and\
  \bibinfo {author} {\bibfnamefont {P.}~\bibnamefont {Hagenmuller}},\ }\href
  {\doibase DOI: 10.1016/0022-3697(78)90200-7} {\bibfield  {journal} {\bibinfo
  {journal} {J. Phys. Chem. Solids}\ }\textbf {\bibinfo {volume} {39}},\
  \bibinfo {pages} {55 } (\bibinfo {year} {1978})}\BibitemShut {NoStop}%
\bibitem [{\citenamefont {Meisenheimer}\ and\ \citenamefont
  {Swalen}(1961)}]{Meisenheimer1961}%
  \BibitemOpen
  \bibfield  {author} {\bibinfo {author} {\bibfnamefont {R.~G.}\ \bibnamefont
  {Meisenheimer}}\ and\ \bibinfo {author} {\bibfnamefont {J.~D.}\ \bibnamefont
  {Swalen}},\ }\href {\doibase 10.1103/PhysRev.123.831} {\bibfield  {journal}
  {\bibinfo  {journal} {Phys. Rev.}\ }\textbf {\bibinfo {volume} {123}},\
  \bibinfo {pages} {831} (\bibinfo {year} {1961})}\BibitemShut {NoStop}%
\bibitem [{\citenamefont {Takatsu}\ \emph {et~al.}(2009)\citenamefont
  {Takatsu}, \citenamefont {Yoshizawa}, \citenamefont {Yonezawa},\ and\
  \citenamefont {Maeno}}]{Takatsu2009}%
  \BibitemOpen
  \bibfield  {author} {\bibinfo {author} {\bibfnamefont {H.}~\bibnamefont
  {Takatsu}}, \bibinfo {author} {\bibfnamefont {H.}~\bibnamefont {Yoshizawa}},
  \bibinfo {author} {\bibfnamefont {S.}~\bibnamefont {Yonezawa}}, \ and\
  \bibinfo {author} {\bibfnamefont {Y.}~\bibnamefont {Maeno}},\ }\href
  {\doibase 10.1103/PhysRevB.79.104424} {\bibfield  {journal} {\bibinfo
  {journal} {Phys. Rev. B}\ }\textbf {\bibinfo {volume} {79}},\ \bibinfo
  {pages} {104424} (\bibinfo {year} {2009})}\BibitemShut {NoStop}%
\bibitem [{\citenamefont {Chapon}\ \emph {et~al.}(2011)\citenamefont {Chapon},
  \citenamefont {Manuel}, \citenamefont {Damay}, \citenamefont {Toledano},
  \citenamefont {Hardy},\ and\ \citenamefont {Martin}}]{Chapon2011}%
  \BibitemOpen
  \bibfield  {author} {\bibinfo {author} {\bibfnamefont {L.~C.}\ \bibnamefont
  {Chapon}}, \bibinfo {author} {\bibfnamefont {P.}~\bibnamefont {Manuel}},
  \bibinfo {author} {\bibfnamefont {F.}~\bibnamefont {Damay}}, \bibinfo
  {author} {\bibfnamefont {P.}~\bibnamefont {Toledano}}, \bibinfo {author}
  {\bibfnamefont {V.}~\bibnamefont {Hardy}}, \ and\ \bibinfo {author}
  {\bibfnamefont {C.}~\bibnamefont {Martin}},\ }\href {\doibase
  10.1103/PhysRevB.83.024409} {\bibfield  {journal} {\bibinfo  {journal} {Phys.
  Rev. B}\ }\textbf {\bibinfo {volume} {83}},\ \bibinfo {pages} {024409}
  (\bibinfo {year} {2011})}\BibitemShut {NoStop}%
\bibitem [{\citenamefont {Toth}\ \emph {et~al.}(2011)\citenamefont {Toth},
  \citenamefont {Lake}, \citenamefont {Kimber}, \citenamefont {Pieper},
  \citenamefont {Reehuis}, \citenamefont {Islam}, \citenamefont {Zaharko},
  \citenamefont {Ritter}, \citenamefont {Hill}, \citenamefont {Ryll},
  \citenamefont {Kiefer}, \citenamefont {Argyriou},\ and\ \citenamefont
  {Williams}}]{Toth2011}%
  \BibitemOpen
  \bibfield  {author} {\bibinfo {author} {\bibfnamefont {S.}~\bibnamefont
  {Toth}}, \bibinfo {author} {\bibfnamefont {B.}~\bibnamefont {Lake}}, \bibinfo
  {author} {\bibfnamefont {S.~A.~J.}\ \bibnamefont {Kimber}}, \bibinfo {author}
  {\bibfnamefont {O.}~\bibnamefont {Pieper}}, \bibinfo {author} {\bibfnamefont
  {M.}~\bibnamefont {Reehuis}}, \bibinfo {author} {\bibfnamefont {A.~T. M.~N.}\
  \bibnamefont {Islam}}, \bibinfo {author} {\bibfnamefont {O.}~\bibnamefont
  {Zaharko}}, \bibinfo {author} {\bibfnamefont {C.}~\bibnamefont {Ritter}},
  \bibinfo {author} {\bibfnamefont {A.~H.}\ \bibnamefont {Hill}}, \bibinfo
  {author} {\bibfnamefont {H.}~\bibnamefont {Ryll}}, \bibinfo {author}
  {\bibfnamefont {K.}~\bibnamefont {Kiefer}}, \bibinfo {author} {\bibfnamefont
  {D.~N.}\ \bibnamefont {Argyriou}}, \ and\ \bibinfo {author} {\bibfnamefont
  {A.~J.}\ \bibnamefont {Williams}},\ }\href {\doibase
  10.1103/PhysRevB.84.054452} {\bibfield  {journal} {\bibinfo  {journal} {Phys.
  Rev. B}\ }\textbf {\bibinfo {volume} {84}},\ \bibinfo {pages} {054452}
  (\bibinfo {year} {2011})}\BibitemShut {NoStop}%
\bibitem [{\citenamefont {Islam}\ and\ \citenamefont
  {Toth}(2012)}]{nazmul3000}%
  \BibitemOpen
  \bibfield  {author} {\bibinfo {author} {\bibfnamefont {N.}~\bibnamefont
  {Islam}}\ and\ \bibinfo {author} {\bibfnamefont {S.}~\bibnamefont {Toth}},\
  }\href@noop {} {} (\bibinfo {year} {2012}),\ \bibinfo {note} {method is to be
  published}\BibitemShut {NoStop}%
\bibitem [{\citenamefont {Abernathy}\ \emph {et~al.}(2012)\citenamefont
  {Abernathy}, \citenamefont {Stone}, \citenamefont {Loguillo}, \citenamefont
  {Lucas}, \citenamefont {Delaire}, \citenamefont {Tang}, \citenamefont {Lin},\
  and\ \citenamefont {Fultz}}]{arcs}%
  \BibitemOpen
  \bibfield  {author} {\bibinfo {author} {\bibfnamefont {D.~L.}\ \bibnamefont
  {Abernathy}}, \bibinfo {author} {\bibfnamefont {M.~B.}\ \bibnamefont
  {Stone}}, \bibinfo {author} {\bibfnamefont {M.~J.}\ \bibnamefont {Loguillo}},
  \bibinfo {author} {\bibfnamefont {M.~S.}\ \bibnamefont {Lucas}}, \bibinfo
  {author} {\bibfnamefont {O.}~\bibnamefont {Delaire}}, \bibinfo {author}
  {\bibfnamefont {X.}~\bibnamefont {Tang}}, \bibinfo {author} {\bibfnamefont
  {J.~Y.~Y.}\ \bibnamefont {Lin}}, \ and\ \bibinfo {author} {\bibfnamefont
  {B.}~\bibnamefont {Fultz}},\ }\href@noop {} {\bibfield  {journal} {\bibinfo
  {journal} {Rev. Sci. Instrum.}\ }\textbf {\bibinfo {volume} {83}},\ \bibinfo
  {pages} {15114} (\bibinfo {year} {2012})}\BibitemShut {NoStop}%
\bibitem [{\citenamefont {Bewley}\ \emph {et~al.}(2006)\citenamefont {Bewley},
  \citenamefont {Eccleston}, \citenamefont {McEwen}, \citenamefont {Hayden},
  \citenamefont {Dove}, \citenamefont {Bennington}, \citenamefont {Treadgold},\
  and\ \citenamefont {Coleman}}]{Bewley06}%
  \BibitemOpen
  \bibfield  {author} {\bibinfo {author} {\bibfnamefont {R.}~\bibnamefont
  {Bewley}}, \bibinfo {author} {\bibfnamefont {R.}~\bibnamefont {Eccleston}},
  \bibinfo {author} {\bibfnamefont {K.}~\bibnamefont {McEwen}}, \bibinfo
  {author} {\bibfnamefont {S.}~\bibnamefont {Hayden}}, \bibinfo {author}
  {\bibfnamefont {M.}~\bibnamefont {Dove}}, \bibinfo {author} {\bibfnamefont
  {S.}~\bibnamefont {Bennington}}, \bibinfo {author} {\bibfnamefont
  {J.}~\bibnamefont {Treadgold}}, \ and\ \bibinfo {author} {\bibfnamefont
  {R.}~\bibnamefont {Coleman}},\ }\href {\doibase 10.1016/j.physb.2006.05.328}
  {\bibfield  {journal} {\bibinfo  {journal} {Physica B}\ }\textbf {\bibinfo
  {volume} {385-386}},\ \bibinfo {pages} {1029 } (\bibinfo {year}
  {2006})}\BibitemShut {NoStop}%
\bibitem [{\citenamefont {Perring}\ \emph {et~al.}(2009)\citenamefont
  {Perring}, \citenamefont {Ewings},\ and\ \citenamefont {Duijn}}]{Horace}%
  \BibitemOpen
  \bibfield  {author} {\bibinfo {author} {\bibfnamefont {T.~G.}\ \bibnamefont
  {Perring}}, \bibinfo {author} {\bibfnamefont {R.~A.}\ \bibnamefont {Ewings}},
  \ and\ \bibinfo {author} {\bibfnamefont {J.~V.}\ \bibnamefont {Duijn}},\
  }\href {http://horace.isis.rl.ac.uk} {\bibfield  {journal} {\bibinfo
  {journal} {unpublished, http://horace.isis.rl.ac.uk}\ } (\bibinfo {year}
  {2009})}\BibitemShut {NoStop}%
\bibitem [{\citenamefont {Tennant}\ and\ \citenamefont
  {McMorrow}(1995)}]{Rescal}%
  \BibitemOpen
  \bibfield  {author} {\bibinfo {author} {\bibfnamefont {A.}~\bibnamefont
  {Tennant}}\ and\ \bibinfo {author} {\bibfnamefont {D.}~\bibnamefont
  {McMorrow}},\ }\href
  {http://www.ill.eu/html/instruments-support/computing-for-science/cs-software/all-software/matlab-ill/rescal-for-matlab/}
  {} (\bibinfo {year} {1995}),\ \bibinfo {note} {unpublished}\BibitemShut
  {NoStop}%
\bibitem [{\citenamefont {Popovici}(1975)}]{popovici75}%
  \BibitemOpen
  \bibfield  {author} {\bibinfo {author} {\bibfnamefont {M.}~\bibnamefont
  {Popovici}},\ }\href {\doibase 10.1107/S0567739475001088} {\bibfield
  {journal} {\bibinfo  {journal} {Acta Cryst. A}\ }\textbf {\bibinfo {volume}
  {31}},\ \bibinfo {pages} {507} (\bibinfo {year} {1975})}\BibitemShut
  {NoStop}%
\bibitem [{\citenamefont {Nagamiya}(1967)}]{Nagamiya67}%
  \BibitemOpen
  \bibfield  {author} {\bibinfo {author} {\bibfnamefont {T.}~\bibnamefont
  {Nagamiya}},\ }\href@noop {} {\emph {\bibinfo {title} {Solid State Phys.}}},\
  edited by\ \bibinfo {editor} {\bibfnamefont {D.~T.}\ \bibnamefont
  {F.~Seits}}\ and\ \bibinfo {editor} {\bibfnamefont {H.}~\bibnamefont
  {Ehrenreich}},\ Vol.~\bibinfo {volume} {20}\ (\bibinfo  {publisher} {Academic
  Press, New York, London},\ \bibinfo {year} {1967})\ p.\ \bibinfo {pages}
  {305}\BibitemShut {NoStop}%
\bibitem [{\citenamefont {White}\ \emph {et~al.}(1965)\citenamefont {White},
  \citenamefont {Sparks},\ and\ \citenamefont {Ortenburger}}]{White1965}%
  \BibitemOpen
  \bibfield  {author} {\bibinfo {author} {\bibfnamefont {R.~M.}\ \bibnamefont
  {White}}, \bibinfo {author} {\bibfnamefont {M.}~\bibnamefont {Sparks}}, \
  and\ \bibinfo {author} {\bibfnamefont {I.}~\bibnamefont {Ortenburger}},\
  }\href {\doibase 10.1103/PhysRev.139.A450} {\bibfield  {journal} {\bibinfo
  {journal} {Phys. Rev.}\ }\textbf {\bibinfo {volume} {139}},\ \bibinfo {pages}
  {A450} (\bibinfo {year} {1965})}\BibitemShut {NoStop}%
\bibitem [{\citenamefont {Coldea}(1997)}]{coldeathesis}%
  \BibitemOpen
  \bibfield  {author} {\bibinfo {author} {\bibfnamefont {R.}~\bibnamefont
  {Coldea}},\ }\emph {\bibinfo {title} {Neutron scattering studies of two
  magnetic phase transitions}},\ \href@noop {} {Ph.D. thesis},\ \bibinfo
  {school} {University of Oxford} (\bibinfo {year} {1997})\BibitemShut
  {NoStop}%
\bibitem [{\citenamefont {Chernyshev}(2009)}]{Chernyshevpc}%
  \BibitemOpen
  \bibfield  {author} {\bibinfo {author} {\bibfnamefont {A.~L.}\ \bibnamefont
  {Chernyshev}},\ }\href@noop {} {\bibfield  {journal} {\bibinfo  {journal}
  {(Private Communication)}\ } (\bibinfo {year} {2009})}\BibitemShut {NoStop}%
\bibitem [{\citenamefont {Fishman}(2011)}]{Fishman2011}%
  \BibitemOpen
  \bibfield  {author} {\bibinfo {author} {\bibfnamefont {R.~S.}\ \bibnamefont
  {Fishman}},\ }\href {\doibase 10.1103/PhysRevB.84.052405} {\bibfield
  {journal} {\bibinfo  {journal} {Phys. Rev. B}\ }\textbf {\bibinfo {volume}
  {84}},\ \bibinfo {pages} {052405} (\bibinfo {year} {2011})}\BibitemShut
  {NoStop}%
\end{thebibliography}%
\end{document}